\documentclass[aps,pra,twocolumn,showpacs]{revtex4-1}
\usepackage{amsmath}
\usepackage[dvips,xdvi]{graphicx}
\usepackage{amssymb}
\usepackage{epsfig}

\begin{document}
\title{Dynamics in spinor condensates controlled by a microwave dressing field}
\author{L. Zhao}
\author{J. Jiang}
\author{T. Tang}
\author{M. Webb}
\author{Y. Liu}
\email{yingmei.liu@okstate.edu} \affiliation{Department of
Physics, Oklahoma State University, Stillwater, OK 74078}
\date{\today}

\begin{abstract}
We experimentally study spin dynamics in a sodium antiferromagnetic
spinor condensate with off-resonant microwave
pulses. In contrast to a magnetic field, a microwave dressing
field enables us to explore rich spin dynamics under the influence
of a negative net quadratic Zeeman shift $q_{\rm net}$. We find an
experimental signature to determine the sign of $q_{\rm net}$, and
observe harmonic spin population oscillations at every $q_{\rm
net}$ except near each separatrix in phase space where spin
oscillation period diverges. In the negative and positive
$q_{\rm net}$ regions, we also observe a
remarkably different relationship between each separatrix and the
magnetization.
Our data confirms an important prediction derived from the 
mean-field theory: spin-mixing dynamics in spin-1 condensates substantially depends
on the sign of the ratio of $q_{\rm net}$ and the spin-dependent interaction energy. This work may thus be the first to use only
one atomic species to reveal mean-field spin dynamics, especially
the separatrix, which are predicted to appear differently in
spin-1 antiferromagnetic and ferromagnetic spinor condensates.
\end{abstract}

\pacs{32.60.+i, 67.85.Hj, 03.75.Kk, 03.75.Mn}

\maketitle

An atomic Bose-Einstein condensate (BEC) is a state where all
atoms have a single collective wavefunction for their spatial
degrees of freedom. The key benefit of spinor BECs is the
additional spin degree of freedom. Together with Feshbach
resonances and optical lattices which tune the interatomic
interactions, spinor BECs constitute a fascinating collective
quantum system offering an unprecedented degree of control over
such parameters as spin, temperature, and the dimensionality of
the system~\cite{StamperKurnRMP, Ueda}. Spinor BECs have
become one of the fastest moving research frontiers in the past
fifteen years. A number of atomic species have proven to be 
perfect candidates in the study of spinor BECs, such as $F$=1 and
$F$=2 hyperfine spin states of $^{87}$Rb
atoms~\cite{Chapman2005,StamperKurnRMP,Ueda,bloch,f2Hirano,f2sengstock1,f2sengstock2},
and $F$=1 hyperfine spin manifolds of $^{23}$Na
atoms~\cite{faraday,fluctuation,Raman2011,black,Gerbier2012}.
Magnetic fields can induce the quadratic Zeeman energy shift
$q_{\rm B}$. Many interesting phenomena driven by an interplay
between $q_{\rm B}$ and the spin-dependent interaction energy $c$ have been
experimentally demonstrated in spinor BECs, such as spin
population dynamics~\cite{f2sengstock2, faraday, Chapman2005,
bloch,StamperKurnRMP,Ueda, black, f2sengstock1,f2Hirano}, quantum
number fluctuation~\cite{You2007,fluctuation}, various quantum
phase
transitions~\cite{StamperKurnRMP,faraday,Raman2011,Gerbier2012},
and quantum spin-nematic squeezing~\cite{Chapman2012}. Such
systems have been successfully described with a classical
two-dimensional phase space~\cite{StamperKurnRMP,Ueda,
You2003,You2005,Lamacraft2011}, a rotor model~\cite{DasSarma2010},
or a quantum model~\cite{You2007,Lamacraft2011}.

\begin{figure}[t]
\includegraphics[width=85mm]{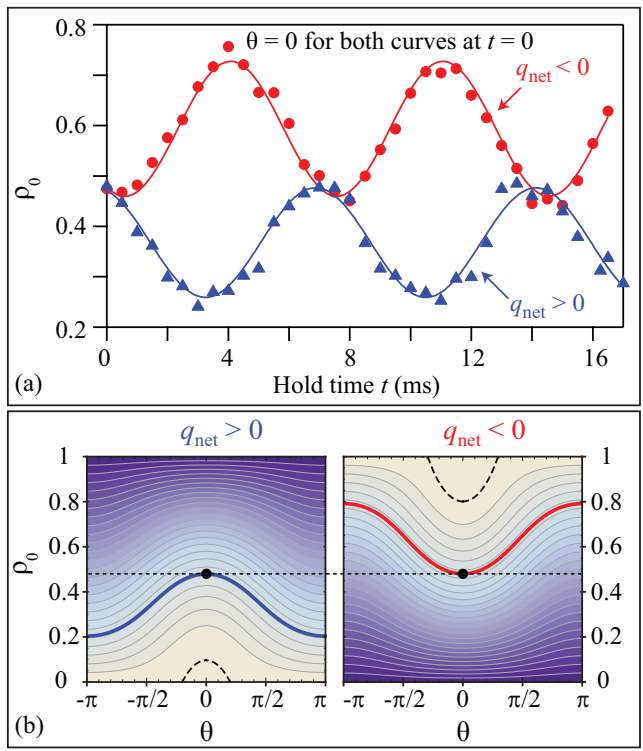}
\caption{(color online)(a). Time evolutions of $\rho_0$ at $q_{\rm
net}/h= + 93~\rm{Hz}>0$ (solid blue triangles) and $q_{\rm
net}/h=-83~\rm{Hz}<0$ (solid red circles) with $m=0$ and
$c/h=52~\rm{Hz}$. It is important to note that the two curves
start from the same initial state with $\theta|_{t=0}=0$. Solid
lines are sinusoidal fits to the data. (b) Equal-energy contour
plots based on Eq.~\ref{eqn:E} for the two experimental conditions
shown in Fig.~\ref{oscillation}(a), i.e., $q_{\rm net}>0$ (left)
and $q_{\rm net}<0$ (right). The heavy solid blue and red lines
represent the energy of the above two experimental conditions,
respectively. The dotted black horizontal line is to emphasize the
fact that the above two experiments start with the same initial
state which is marked by the solid black circles. Dashed black
lines represent the energy of the separatrix between the running
and oscillatory phase solutions. Darker colors correspond to lower
energies.} \label{oscillation}
\end{figure}

\begin{figure}[t]
\includegraphics[width=85mm]{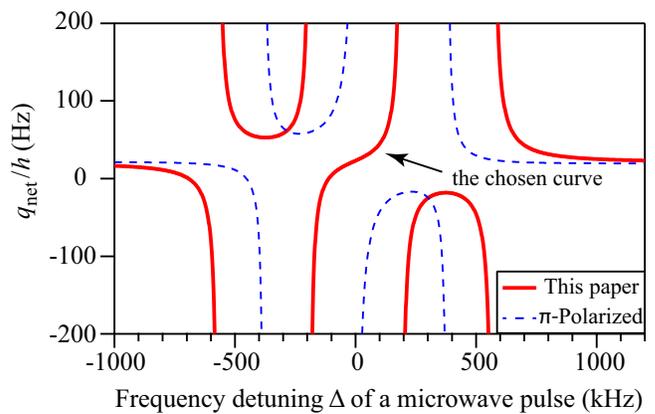}
\caption{(color online) $q_{\rm net}$ as a function of $\Delta$.
The residual magnetic field is $B$=270~mG. Dashed blue lines and
solid red lines represent the predictions derived from
Eq.~\ref{eqn:dressing} when the microwave pulse is purely
$\pi$-polarized and when the pulse has a specially-chosen
polarization, respectively (see text). In this paper, $\Delta$ is
tuned within the range of $-190$~kHz to $190$~kHz.} \label{uwave}
\end{figure}

\begin{figure*}[t]
\includegraphics[width=175mm]{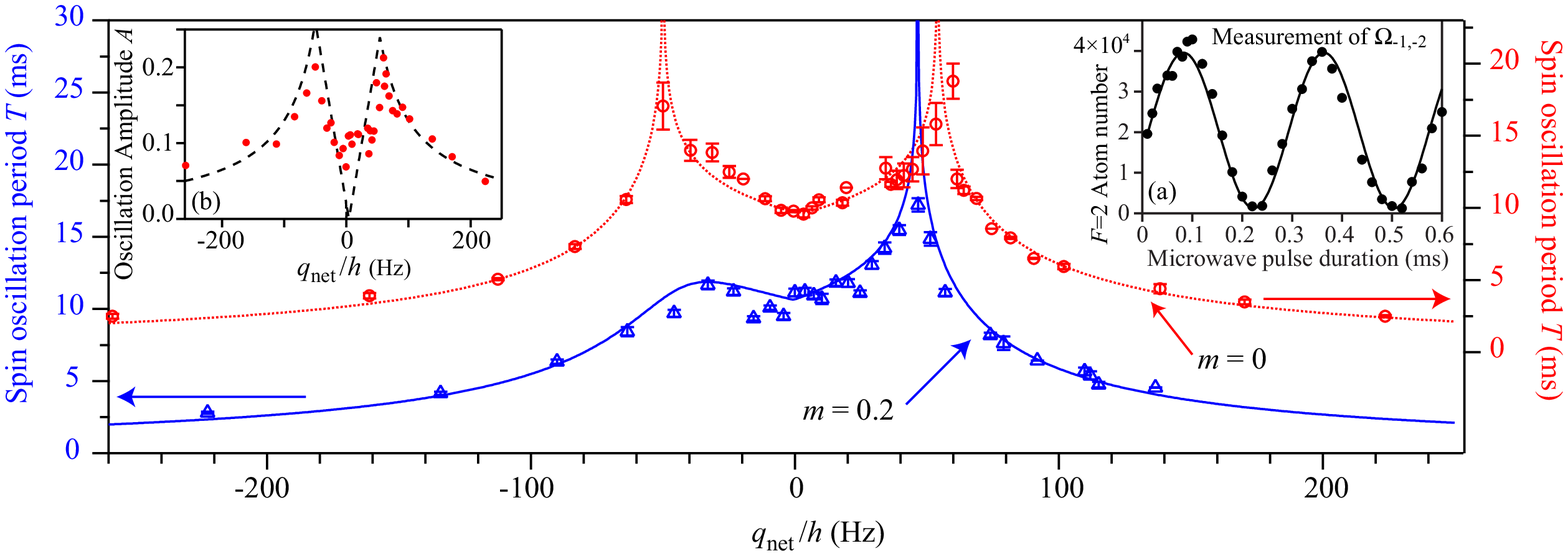}
\caption{(color online) The spin oscillation period as a function
of $q_{\rm net}$ for $m=0$ (open red circles) and $m=0.2$ (open
blue triangles). The lines are fits based on Eq.~\ref{eqn:E},
which yield the fit parameters: $\rho_0|_{t=0}=0.48$,
$\theta|_{t=0}=0$, and $c/h=52~\rm{Hz}$ for $m=0$; and
$\rho_0|_{t=0}=0.48$, $\theta|_{t=0}=0$, and $c/h=47~\rm{Hz}$ for
$m=0.2$. The fit parameters are within the 5\% uncertainty of our
measurements. Note the different scales of the left and right
vertical axes. Inset (a): the number of $F=2$ atoms excited by a
resonant microwave pulse as a
function of the pulse duration. The solid line is a sinusoidal fit
to extract the on-resonance Rabi frequency
$\Omega_{-1,-2}$. Inset (b): amplitudes $A$ of spin
oscillations shown in the main figure as a function of $q_{\rm
net}$ at $m=0$. The dashed black line is a fit based
on Eq.~\ref{eqn:E} with the same set of fit parameters as that
applied in the main figure.} \label{period}
\end{figure*}

In this paper, we experimentally study spin-mixing dynamics in a
$F$=1 sodium spinor condensate starting from a nonequilibrium
initial state, as a result of antiferromagnetic spin-dependent
interactions and the quadratic Zeeman energy $q_{\rm M}$ induced
by an off-resonant microwave pulse. In contrast to a magnetic
field, a microwave dressing field enables us to explore rich spin
dynamics under the influence of a negative net quadratic Zeeman
energy shift $q_{\rm net}$. A method to
characterize the microwave dressing field is also explained. In both
negative and positive $q_{\rm net}$ regions, we observe spin
population oscillations resulted from coherent collisional
interconversion among two $|F=1, m_F =0 \rangle$ atoms, one $|F=1,
m_F =+1 \rangle$ atom, and one $|F=1, m_F =-1 \rangle$ atom. In every spin oscillation studied in this paper, our
data shows that the population of the $m_F=0$ state averaged over time
is always larger (or
smaller) than its initial value as long as $q_{\rm net}<0$ (or
$q_{\rm net}>0$). This observation provides an experimental signature to
determine the sign of $q_{\rm net}$. We also find a remarkably
different relationship between the total magnetization $m$ and a
separatrix in phase space where spin oscillation period diverges:
the position of the separatrix moves slightly with $m$ in the
positive $q_{\rm net}$ region, while the separatrix quickly
disappears when $m$ is away from zero in the negative $q_{\rm
net}$ region. Our data confirms an important prediction derived
by Ref.~\cite{Lamacraft2011}: the spin-mixing dynamics in $F$=1 spinor condensates substantially depends
on the sign of $R=q_{\rm net}/c$. This work may thus be the first to use only one atomic
species to reveal mean-field spin dynamics, especially the
separatrix, which are predicted to appear differently
in $F$=1 antiferromagnetic and ferromagnetic spinor condensates.

Similar to Ref.~\cite{StamperKurnRMP,You2005}, we take into account the
conservation of $m$ and the total atom number. Because no spin domains and spatial
modes are observed in our system, the single
spatial mode approximation (SMA), in which all spin states have the
same spatial wavefunction, appears to be a proper
theoretical model to understand our data. Spin-mixing dynamics in
a $F$=1 spinor BEC can thus be described with a two-dimensional
($\rho_0$ vs $\theta$) phase space, where the fractional population
$\rho_{\rm m_F}$ and the phase $\theta_{\rm m_F}$ of each $m_F$
state are independent of position. The BEC energy $E$ and the
time evolution of $\rho_0$ and $\theta$ may be expressed
as~\cite{StamperKurnRMP,You2005}
\begin{eqnarray}  \label{eqn:E}
&E=q_{\rm net}(1-\rho_0)+c
\rho_0[(1-\rho_0)+\sqrt{(1-\rho_0)^2-m^2}\cos\theta], \nonumber\\
&\dot{\rho_0}=-(2/\hbar)\partial E/\partial \theta,
~~\dot{\theta}=(2/\hbar)\partial E/\partial \rho_0~.
\end{eqnarray}
Here $q_{\rm net}=q_{\rm B}+q_{\rm M}$, $\theta=\theta_{+1} + \theta_{-1} -2\theta_0$ is the relative phase among the three $m_{\rm F}$ spin states, and $\hbar$ is the reduced
Planck constant. The induced linear Zeeman shift remains the same
during the collisional spin interconversion and is thus ignored.
The spin-dependent
interaction energy is $c=c_2 \langle n\rangle$, where $\langle
n\rangle$ is the mean BEC density and $c_2$ is the spin dependent
interaction coefficient. The total magnetization is
$m=\rho_{+1}-\rho_{-1}$. It is well known that $q_B\propto B^2>0$,
and $c_2>0$ (or $c_2<0$) in $F$=1 $^{23}$Na (or $^{87}$Rb) spinor
BECs. Spin-dynamics in $F$=1 antiferromagnetic and ferromagnetic
spinor BECs have been studied in magnetic fields where $q_{\rm
net}>0$ with $^{23}$Na and $^{87}$Rb atoms,
respectively~\cite{StamperKurnRMP}. A few methods have been
explored for generating a negative quadratic Zeeman shift, such as
via a microwave dressing
field~\cite{uwaveBloch,Stamper-Kurn2009,StamperKurnRMP,Raman2011,StamperKurnPhD}
or through a linearly polarized off-resonant laser
beam~\cite{Pfau2007}. In this paper, we choose the first method.

The experimental setup is similar to that illustrated in our
previous work~\cite{JiangBEC}. Hot $^{23}$Na atoms are
slowed by a spin-flip Zeeman slower, captured in a standard
magneto-optical trap, cooled through a polarization gradient
cooling process to 40 $\mu$K, and loaded into a crossed optical
dipole trap originating from a linearly-polarized high power IR
laser at 1064~nm. After an optimized 6~s forced evaporative
cooling process, a pure $F$=1 BEC of $1\times10^5$ sodium atoms is
created. We can polarize atoms in the $F$=1 BEC fully to the
$|F=1, m_F =-1 \rangle$ state by applying a weak magnetic field
gradient during the first half of the forced evaporation (or fully
to the $|F=1, m_F =0 \rangle$ state by adding a very strong
magnetic bias field during the entire 6~s forced evaporation). We
then ramp up a small magnetic bias field with its strength $B$
being 270~mG, while turning off the field gradient. An rf-pulse
resonant with the linear Zeeman splitting is applied to prepare an
initial state with any desired combination of the three $m_F$
states. To generate sufficiently large $q_{\rm net}$ with
off-resonant microwave pulses, a microwave antenna designed for a
frequency near the $|F=1\rangle\leftrightarrow|F=2\rangle$
transition is placed a few inches above the center of the
magneto-optical trap, and connected to a function generator
outputting a maximum radiation power of 10~W. After various hold
time $t$ in the optical dipole trap, populations of the multiple
spin states are then measured via the standard absorption imaging
preceded by a 3~ms Stern-Gerlach separation and a 7~ms time of
flight.

We observe spin oscillations at every given value of $q_{\rm net}$
within a wide range, i.e., $-240~{\rm{Hz}}~\leq~q_{\rm
net}/h~\leq~240~\rm{Hz}$. Here $h$ is the Planck constant. Typical
time evolutions of $\rho_0$ starting with the same nonequilibrium
initial state at a negative and a positive $q_{\rm net}$ are shown
in Fig.~\ref{oscillation}(a). We find that these evolutions can be
well fit by sinusoidal functions of the similar oscillation period
$T$ and amplitude $A$. On the other hand, our data in
Fig.~\ref{oscillation}(a) shows that the value of $\langle
\rho_0\rangle$ is drastically different in the two spin
oscillations: $\langle \rho_0\rangle>\rho_0|_{t=0}$ as long as
$q_{\rm net}<0$, while $\langle \rho_0\rangle<\rho_0|_{t=0}$ if
$q_{\rm net}>0$. Here $\langle \rho_0\rangle$ is the average value
of $\rho_0$ over time in a spin oscillation and $\rho_0|_{t=0}$ is the
initial value of $\rho_0$. This phenomenon is observed at every
value of $q_{\rm net}$ when spin oscillations start with the same
initial state, although the period $T$ and amplitude $A$ change
with $q_{\rm net}$. The above observations agree well with
predictions from the mean-field SMA theory (i.e., Eq.~\ref{eqn:E})
as shown by the heavy solid lines in Fig.~\ref{oscillation}(b):
$\rho_0$ is limited between $(\rho_0|_{t=0}-2A)$ and
$\rho_0|_{t=0}$ at $q_{\rm net}>0$, while it is restricted between
$\rho_0|_{t=0}$ and $(\rho_0|_{t=0}+2A)$ at $q_{\rm net}<0$. We
can thus use the phenomenon to conveniently determine the sign of
$q_{\rm net}$, i.e., by comparing the value of $\langle
\rho_0\rangle$ of a spin oscillation to the value of
$\rho_0|_{t=0}$.

On the other hand, the exact value of $q_{\rm net}$ is carefully
calibrated based on Eq.~\ref{eqn:dressing} with a few experimental
parameters, such as the polarization and frequency of a microwave
pulse. Every microwave pulse used in this paper is detuned by
$\Delta$ from the $| F=1, m_F = 0 \rangle \leftrightarrow | F=2,
m_F = 0 \rangle$ transition. A purely $\pi$-polarized microwave
pulse has been a popular choice in some
publications~\cite{uwaveBloch,Stamper-Kurn2009,StamperKurnRMP,Raman2011,StamperKurnPhD}.
However, the microwave pulse used in this paper has a
specially-chosen polarization in order to easily access every
positive and negative value of $q_{\rm net}$ just by continuously
tuning $\Delta$, as shown in Fig.~\ref{uwave}. We define $k$ as 0
or $\pm 1$ for a $\pi$ or a $\sigma_{\pm}$ polarized microwave
pulse, respectively. For a given polarization $k$, the allowed transition is
$|F=1, m_F\rangle\leftrightarrow|F=2, m_{F}+k \rangle$ and
its on-resonance Rabi frequency is $\Omega_{m_F,m_{F}+k}\propto
\sqrt{I_k}C_{m_F,m_{F}+k}$, where $C_{m_F,m_{F}+k}$ is the
Clebsch-Gordan coefficient of the transition and $I_k$ is the intensity of this purely polarized microwave pulse. We also define
$\Delta_{m_F,m_{F}+k}=\Delta-[(m_{F}+k)/2-(-m_F/2)]\mu_BB$ as the
frequency detuning of the microwave pulse with respect to the
$|F=1, m_F\rangle\rightarrow|F=2, m_{F}\rangle$ transition, where
$\mu_B$ is the Bohr magneton. Similar to
Refs.~\cite{uwaveBloch,StamperKurnPhD}, we express the value of
$q_{\rm net}$ as
\begin{align}  \label{eqn:dressing}
&q_{\rm net}=q_{\rm B}+q_{\rm M} \nonumber\\
&~~~~~=~aB^2h+(\delta E|_{m_F=1}+\delta E|_{m_F=-1}-2 \delta E|_{m_F=0})/2,\nonumber\\
&\delta E|_{m_F}=\frac{h}{4}\sum_{k}
\frac{\Omega_{m_F,m_{F}+k}^2}{\Delta_{m_F,m_{F}+k}}~,
\end{align}
where $a \approx 277~\rm{Hz/G^2}$ (or $a \approx 71~\rm{Hz/G^2}$)
for $F$=1 $^{23}$Na (or $^{87}$Rb) atoms. Due to the limited power
of our microwave function generator, we obtain a desired value of
$q_{\rm net}$ by choosing a proper $\Delta$ within the range of
$-190~\rm kHz$ to $190~\rm kHz$ at a fixed intensity, as
shown in Fig.~\ref{uwave}. The on-resonance Rabi frequencies of
our microwave pulses are $\Omega_{-1,-2}=3.6~\rm kHz$,
$\Omega_{0,-1}=2.1~\rm kHz$, $\Omega_{1,0}=1.5~\rm kHz$,
$\Omega_{-1,-1}=\Omega_{0,0}=\Omega_{1,1}=0$,
$\Omega_{-1,0}=1.6~\rm kHz$, $\Omega_{0,1}=2.8~\rm kHz$, and
$\Omega_{1,2}=4.0~\rm kHz$. Another advantage of choosing such
microwave pulses is to conveniently place the microwave antenna on
our apparatus without blocking optical components. In order to
ensure an accurate calibration of $q_{\rm net}$ based on
Eq.~\ref{eqn:dressing}, we measure $\Omega_{m_F,m_{F}+k}$ everyday
by monitoring the number of atoms excited by a resonant microwave
pulse to the $F$=2 state as a function of the pulse duration. A
typical example of the Rabi frequency measurement is shown in the
inset (a) in Fig.~\ref{period}.

The time evolution of $\rho_0$ is fit by a sinusoid to extract the
spin oscillation period $T$ and amplitude $A$ at a given $q_{\rm net}$, as shown in
Fig.~\ref{oscillation}(a). The value of $T$ as a function of
$q_{\rm net}$ is then plotted in Fig.~\ref{period} for $m=0$ and
$m=0.2$, which demonstrates two interesting results. First, when
$m=0$, the spin oscillation is harmonic except near the critical
values (i.e., $q_{\rm net}/h=\pm~52~\rm{Hz}$) where the period
diverges. This agrees with the predictions derived from
Eq.~\ref{eqn:E}, as shown by the dotted red line in
Fig.~\ref{period}. The energy contour $E_{\rm sep}$ where the
oscillation becomes anharmonic is defined as a separatrix in phase
space. A point on the separatrix satisfies the equation
$\dot{\rho_0}=\dot{\theta}=0$ according to the mean-field SMA
theory. In fact for our sodium system with $c>0$, $E_{\rm sep}=
q_{\rm net}$ for $q_{\rm net}>0$, while $E_{\rm sep}= 0$ at $m=0$ for $q_{\rm
net}<0$. Figure~\ref{period} shows that the $T$ vs $q_{\rm net}$
curve is symmetric with respect to the $q_{\rm net}=0$ axis at
$m=0$. The period $T$ decreases rapidly with increasing $|q_{\rm
net}|$ when $|q_{\rm net}|$ is large, which corresponds to the
``Zeeman regime" with running phase solutions. In the opposite
limit, the period only weakly depends on $|q_{\rm net}|$, which
represents the ``interaction regime" with oscillatory phase
solutions. Here $|q_{\rm net}|$ is the absolute value of $q_{\rm
net}$. The value of $\theta$ is (or is not) restricted in the
regions with oscillatory (or running) phase solutions.
Refs~\cite{black, faraday} reported observations of the similar
phenomena for $q_{\rm net}>0$ with a $F$=1 antiferromagnetic
spinor condensate, however, they did not access the negative
$q_{\rm net}$ region.

Figure~\ref{period} also demonstrates a remarkably different
relationship between the total magnetization $m$ and the
separatrix in phase space: the position of the separatrix moves
slightly with $m$ in the positive $q_{\rm net}$ region, while the
separatrix quickly disappears when $m$ is away from zero in the
negative $q_{\rm net}$ region. Good agreements between our data
and the mean-field SMA theory are
shown in the inset (b) and the main figure in Fig.~\ref{period}.
Interestingly, we find that the spin dynamics which appear in our
antiferromagnetic sodium system in the negative $q_{\rm net}$
region exactly mimic what is predicted to occur in a ferromagnetic
spinor condensate in the positive $q_{\rm net}$
region~\cite{Lamacraft2011, You2005}. Note that $R=q_{\rm net}/c$
is negative in both of these two cases. This observation
confirms an important prediction made by
Ref.~\cite{Lamacraft2011}: the spin-mixing dynamics in $F$=1 spinor condensates substantially depends
on the sign of $R$. As a matter of fact, our results in the negative $q_{\rm
net}$ region are similar to those reported with a
$F$=1 ferromagnetic $^{87}$Rb spinor condensate in magnetic fields
where $q_{\rm net}>0$~\cite{Chapman2005,StamperKurnRMP}. Although
the relationship between the separatrix and $m$ in the
ferromagnetic Rb system has not been experimentally explored yet,
our data in Fig.~\ref{period} can be extrapolated to understand
this relationship. This paper may thus be the first to use only
one atomic species to reveal mean-field spin dynamics, especially
the separatrix, which are predicted to appear differently in $F$=1
antiferromagnetic and ferromagnetic spinor condensates.

In conclusion, we have experimentally studied spin dynamics in a
sodium spinor condensate controlled by a microwave dressing field. In both
negative and positive $q_{\rm net}$ regions, we have observed
harmonic spin oscillations and found that the sign of $q_{\rm
net}$ can be determined by comparing $\langle \rho_0 \rangle$ to
$\rho_0|_{t=0}$. Our data also demonstrates that the position
of the separatrix in phase space moves slightly with $m$ in the
positive $q_{\rm net}$ region, while the separatrix quickly
disappears when $m$ is away from zero in the negative $q_{\rm
net}$ region. Our data confirms that the spin-mixing dynamics in $F$=1 spinor condensates substantially depends
on the sign of $R=q_{\rm net}/c$. This paper may thus be the first to use only one atomic
species to reveal mean-field spin dynamics and the separatrix,
which are predicted to appear differently in $F$=1
antiferromagnetic and ferromagnetic spinor condensates. In
addition, the microwave dressing field is able to completely
cancel out $q_{\rm B}$ induced by ambient stray magnetic fields.
This allows us to study interesting but unexplored phenomena at
$q_{\rm net}=0$, for example, realizing a maximally entangled
Dicke state with sodium spinor condensates~\cite{Duan2013}.

We thank the Army Research Office, Oklahoma Center for the
Advancement of Science and Technology, and Oak Ridge Associated
Universities for financial support. M.W. thanks the Niblack Research
Scholar program.

\end{document}